\documentclass[10pt,aps,prb,twocolumn,groupedaddress]{revtex4-1}

\usepackage{natbib}
\usepackage[dvips]{graphicx}
\usepackage{bm}
\usepackage{amsmath, amssymb}

\newcommand{\BFA}{BaFe$_{2}$As$_{2}$}
\newcommand{\BFCA}{Ba(Fe$_{1-x}$Co$_{x}$)$_{2}$As$_{2}$}

\newcommand{\BFCAb}{Ba(Fe$_{0.94}$Co$_{0.06}$)$_{2}$As$_{2}$}

\newcommand{\TN}{$T_{\mathrm{N}}$}
\newcommand{\Tc}{$T_{\mathrm{c}}$}
\newcommand{\cm}{cm$^{-1}$}
\newcommand{\Neff}{$N_{\mathrm{eff}}$}

\begin{document}


\title{Evolution of the optical spectrum with doping in iron-pnictides \BFCA{}}



\author{M.~Nakajima,$^{1,2,3}$ S.~Ishida,$^{1,2,3}$ K.~Kihou,$^{2,3}$ Y.~Tomioka,$^{2,3}$ T.~Ito,$^{2,3}$ Y.~Yoshida,$^{2}$ C.~H.~Lee,$^{2,3}$ H.~Kito,$^{2,3}$ A.~Iyo,$^{2,3}$ H.~Eisaki,$^{2,3}$ K.~M.~Kojima,$^{1,3}$ and S.~Uchida$^{1,3}$}
\affiliation{$^{1}$Department of Physics, University of Tokyo, Tokyo 113-0033, Japan\\
$^{2}$National Institute of Advanced Industrial Science and Technology, Tsukuba 305-8568, Japan\\
$^{3}$JST, Transformative Research-Project on Iron Pnictides (TRIP), Tokyo 102-0075, Japan}


\date{\today}

\begin{abstract}

We investigated the optical spectrum of \BFCA{} single crystals with various doping levels. It is found that the low-energy optical conductivity spectrum of this system can be decomposed into two components: a sharp Drude term and a broad ``incoherent'' term. For the compounds showing magnetic order, a gap appears predominantly in the ``incoherent'' component, while an $s$-wave like superconducting gap opens in both components for highly doped compounds. The Drude weight steadily increases as doping proceeds, consistent with electron doping in this system. On the other hand, the ``incoherent'' spectral weight is almost doping independent, but its spectral feature is intimately connected with the magnetism. We demonstrate that the presence of two distinct components in the optical spectrum well explains the doping and temperature dependences of the dc resistivity.

\end{abstract}


\maketitle

\section{Introduction}

The phase diagram of Fe-based superconductors, specifically of Fe arsenides, is similar to that of the high-\Tc{} cuprates in that the superconducting (SC) phase emerges from the magnetically ordered state by ``doping.'' \cite{Zhao2008, Chen2008, Nandi2009} A difference between the two systems is that the former is the multiband system in contrast to the latter with basically single-band nature. All the five Fe 3$d$ orbitals contribute to the construction of Fermi surfaces, resulting in the presence of multiple Fermi-surface pockets, \cite{Mazin2008, Kuroki2008} as confirmed by angle-resolved photoemission spectroscopy (ARPES). \cite{Hsieh2008, Yi2009} In addition, while the undoped cuprate is an antiferromagnetic Mott insulator, the parent compound of Fe arsenides is metallic, and shows a magnetic or spin-density-wave (SDW) order below the magnetic/structural transition temperature \TN{} accompanied with a reconstruction of Fermi surfaces as observed by ARPES measurements. \cite{Hsieh2008, Yi2009, Shimojima2009}

``Doping'' suppresses the magnetic order and induces superconductivity. \cite{Zhao2008, Chen2008, Nandi2009} Unlike the cuprates, ``doping'' does not necessarily mean chemical tuning of the carrier density. The parent compound also becomes a superconductor upon application of pressure \cite{Kotegawa2009} and even upon substitution of As by isoelectronic P. \cite{Kasahara2009} Therefore, in order to gain insight into the electronic states that give birth to high-\Tc{} superconductivity in FeAs-based compounds, it is crucial to elucidate how the ``doping'' reorganizes the low-lying electronic states of the magnetically ordered phase in the SC regime. This could be done by using advanced spectroscopic methods such as ARPES measurements. Unfortunately, no result of systematic ARPES study has been reported at the moment partly because the sample surface is unstable against reconstruction.

Optical spectrum is a bulk-sensitive and useful energy-resolved probe for investigating doping evolution of charge excitations and dynamics of relevant carriers. Application to the undoped compound revealed an opening of a partial gap in the magnetically ordered state. \cite{Hu2008, Wu2009} It is interesting to explore how this gap evolves with doping.

The optical spectrum can in principle probe a SC gap. Concerning the SC gap in iron-pnictide superconductors, possible symmetry is, as theoretically argued, \cite{Mazin2008, Kuroki2008} $s_{\pm}$ wave which is basically $s$-symmetry with a full gap opening on every part of Fermi surfaces but with the order parameter changing sign across separate Fermi-surface sheets. A previous optical study of K-doped \BFA{} reported a fairly wide gap opening in the in-plane optical spectrum below the SC transition temperature \Tc{}. \cite{Li2008}

In this paper, we choose a system of Co-doped \BFA{}, \BFCA{}, for which measurement of doping evolution of the optical spectrum is possible on single-crystalline samples. To the best of our knowledge, this is the first systematic study of the optical spectroscopy on iron pnictides. We demonstrate that the low-energy conductivity spectrum can be decomposed into a Drude and an ``incoherent'' component, and show the evolution of each component with doping and how a gap (gaps) opens in each component upon the onset of magnetic and superconducting order.

\section{Experimental}

Single crystals of \BFCA{} (nominal Co content $x$ = 0, 0.04, 0.08, and 0.11) were grown using a self-flux method. FeAs and CoAs precursors were prepared from Fe (or Co) and As at 900 $^\circ$C for 10 hours in an evacuated atmosphere. Ba, FeAs, and CoAs were mixed in the atomic ratio 1:$4(1-x)$:$4x$, placed in an Al$_2$O$_3$ crucible, and sealed in a quartz tube. The tube was heated to 1150 $^\circ$C, kept the temperature for 10 h, and cooled to 1050 $^\circ$C for 250 h, followed by decanting the flux. The Co compositions of the samples were determined by inductively coupled plasma (ICP) analysis. It is found that the actual Co contents of the compounds are $x$ = 0.04(1), 0.06(2), and 0.08(4), respectively. Hereafter, we use these ICP-determined compositions. The compounds with $x$ = 0 and 0.04 show a magnetic transition at \TN{} = 138 and 80 K, and those with $x$ = 0.06 and 0.08 undergo a SC transition at \Tc{} = 25 and 20 K, respectively. \Tc{} was defined as the onset of the zero-field-cooled diamagnetic susceptibility, which coincides with the temperature of zero resistivity. These temperatures are close to that reported by other groups. \cite{Ni2008, Chu2009}

A standard four-terminal technique was used for the resistivity measurement. For each $x$, the measurement was made on several crystals from the same batch. The magnitude of resistivity coincides within 5 \% among the crystals with the same $x$, evidencing homogeneity in the Co content. 

The optical reflectivity $R(\omega)$ was measured on the $ab$-plane of the cleaved samples in the frequency range 50-40000 \cm{} at various temperatures by a Fourier transform infrared spectrometer (Bruker IFS113v) and a grating monochromator (JASCO CT-25C). The optical conductivity $\sigma_1(\omega)$ was derived from the Kramers-Kronig transformation of $R(\omega)$. Since the measurement is made in the limited energy region, proper extrapolations are necessary. The Hagen-Rubens or Drude-Lorentz formula is used for the low-energy extrapolation in order to smoothly connect to the spectrum in the measured region and to fit the measured resistivity value at $\omega$ = 0.

\section{Results and discussions}

\subsection{Temperature dependence of resistivity for \BFCA{} and optical response of the undoped compound}

\begin{figure}
\includegraphics[width=80mm,clip]{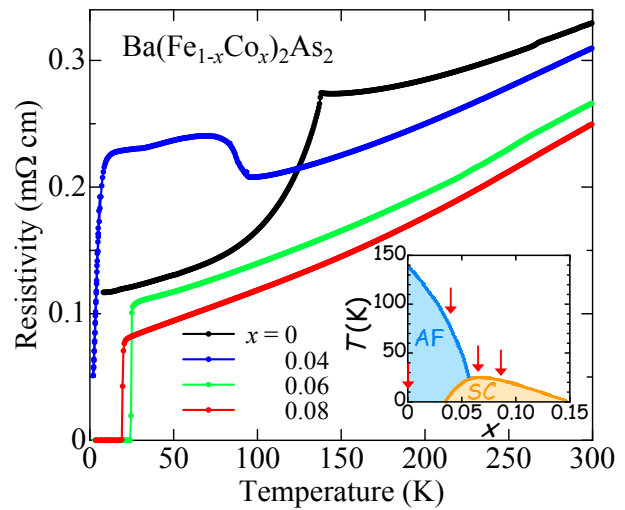}%
\caption{(Color online) Temperature dependence of the in-plane resistivity for \BFCA{} samples ($x$ = 0, 0.04, 0.06, and 0.08). A kink for $x$ = 0 corresponds to a magnetic transition at \TN{} = 138 K. For $x$ = 0.04, \TN{} shifts down to $\sim$ 80 K. The compounds with $x$ = 0.06 and 0.08 show a sharp superconducting transition at \Tc{} = 25 and 20 K, respectively. The inset shows the schematic phase diagram of Co-doped \BFA{}. Red arrows indicate the compositions for which we study in the present work.\label{01}}
\end{figure}

Figure \ref{01} shows temperature dependence of the in-plane resistivity $\rho(T)$ for \BFCA{}. The compositions are indicated in the phase diagram as red arrows (see the inset). Our data are in good agreement with previous reports. \cite{Albenque2009, Fang2009} Resistivity of the undoped compound shows a sharp drop at \TN{} = 138 K and a metallic behavior below \TN{}. For the lightly doped compound ($x$ = 0.04), the resistivity jumps up at 90 K (probably corresponding to a structural transition), and saturates at $\sim$ 80 K which is the reported temperature \TN{} of the magnetic order. \cite{Nandi2009} A spurious resistivity drop is seen below 10 K, but superconductivity is not bulk in this compound. For $x$ = 0.06 and 0.08, there is no indication of the structural/magnetic transition, and instead the resistivity shows a sharp SC transition at 25 and 20 K, respectively. One can see that the overall magnitude of resistivity decreases with doping, but does not strongly depend on $x$. A fairly large residual resistivity component exists for all the samples, which would arise partly from disorder of dopant Co atoms existing in the Fe planes.

\begin{figure}
\includegraphics[width=80mm,clip]{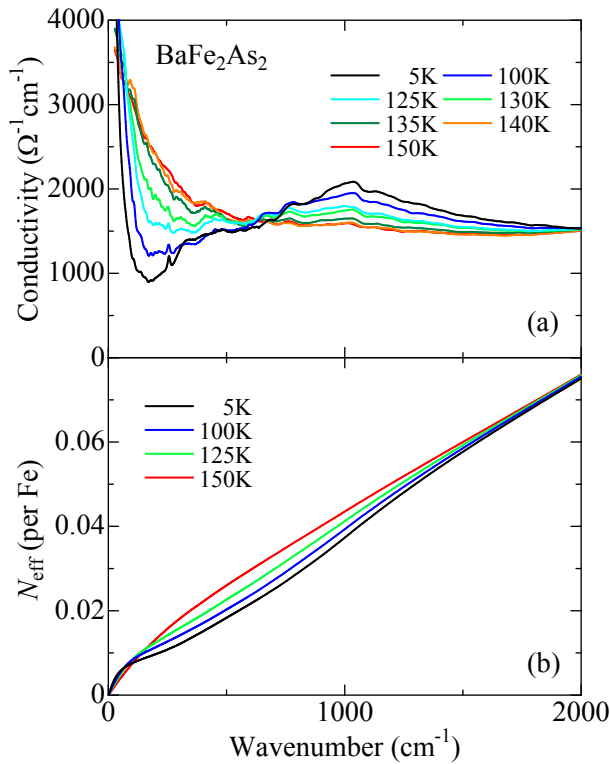}%
\caption{(Color online) (a) Optical conductivity spectrum and (b) \Neff{}($\omega$) of undoped \BFA{} at several temperatures below 150 K. Below \TN{} = 138 K, low-energy conductivity is suppressed, and the suppressed weight is transferred to the higher-energy region up to 2000 \cm{} as evidenced by \Neff{} curves which merge at around 2000 \cm{}. An isosbestic point in $\sigma_1 (\omega)$ is identified at $\omega$ = 650 \cm{}. \label{02}}
\end{figure}

We show temperature evolution of the optical conductivity spectrum $\sigma_1(\omega)$ and its conductivity sum \Neff$(\omega)$ of undoped \BFA{} (Figs.\ \ref{02}(a) and \ref{02}(b));
\begin{equation}
N_{\mathrm{eff}} (\omega) = \frac{2 m_0 V}{\pi e^2} \int^{\omega}_{0} \sigma_1(\omega ^\prime) \mathrm{d} \omega ^\prime,
\end{equation}
where $m_0$ stands for the free electron mass, and $V$ the cell volume containing one Fe atom. The spectrum at $T$ = 150 K just above \TN{} shows slowly decaying conductivity with a peak at $\omega$ = 0 and a long tail extending to 2000 \cm{} or higher. Below \TN{}, the conductivity below 650 \cm{} is suppressed, and most of the suppressed spectral weight is transferred to energies higher than 650 \cm{} forming a peak around 1000 \cm{} as evidenced by the \Neff($\omega$) curves (Fig.\ \ref{02}(b)) merging toward 2000 \cm{}. The isosbestic behavior seen in the doping evolution of the optical conductivity spectrum is a hallmark of strong electronic correlation in the case of cuprates and other transition-metal oxides. It is indicative of a radical reconstruction of the electronic structure involving electronic states over a large energy range, which would also be the case with \BFA{} induced by the onset of the magnetic order. This is in contrast to the gap opening in ordinary CDW/SDW system and also to the well-known Cr metal showing SDW \cite{Boekelheide2007} in which the gap width gradually increases from zero as temperature is lowered below the ordering temperature. Different from these cases, the magnetic transition in the present system might be of first order, \cite{Dong2008} probably because a structural transition always occurs accompanied with the magnetic transition. Note that the gap-energy scale $E_{\mathrm{m}}$, a peak energy of $\sim$ 1000 \cm{}, is huge as compared with the temperature scale \TN{} = 138 K ($E_{\mathrm{m}} / k_{\mathrm{B}} T_{\mathrm{N}} \sim$ 10). It is not trivial if such reconstruction of electronic states and large gap-energy scale are compatible with itinerant spin picture.

\subsection{Evolution of optical spectrum with doping}

\begin{figure}
\includegraphics[width=80mm,bb=127 82 495 755]{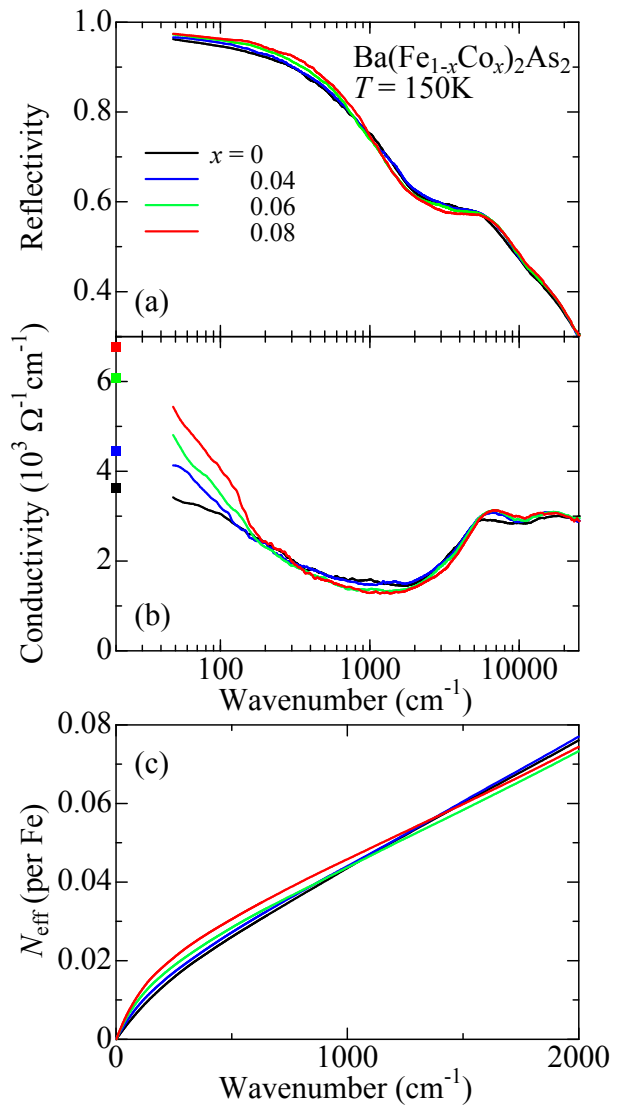}%
\caption{(Color online) (a) Reflectivity spectra, (b) optical conductivity spectra, and (c) conductivity sum \Neff{}($\omega$) at $T$ = 150 K for various doping levels of \BFCA{}. Solid squares in (b) indicate the values of the dc ($\omega$ = 0) conductivity for each composition. \label{03}}
\end{figure}

The reflectivity and optical conductivity spectra at $T$ = 150 K are displayed in Figs.\ \ref{03}(a) and(b) for four Co compositions. This is to see how the Co-doping affects the spectrum of the parent compound in the normal (nonmagnetic) state. In order to see the doping evolution of the spectrum more quantitatively, we calculate \Neff($\omega$), shown in Fig.\ \ref{03}(c). The effect of doping is very weak, only slightly increasing conductivity in the lowest-energy region. The weak doping evolution is what is expected from the band picture in which Co substitution for Fe adds an electron, and only shifts the chemical potential.

\begin{figure}
\includegraphics[width=80mm,clip]{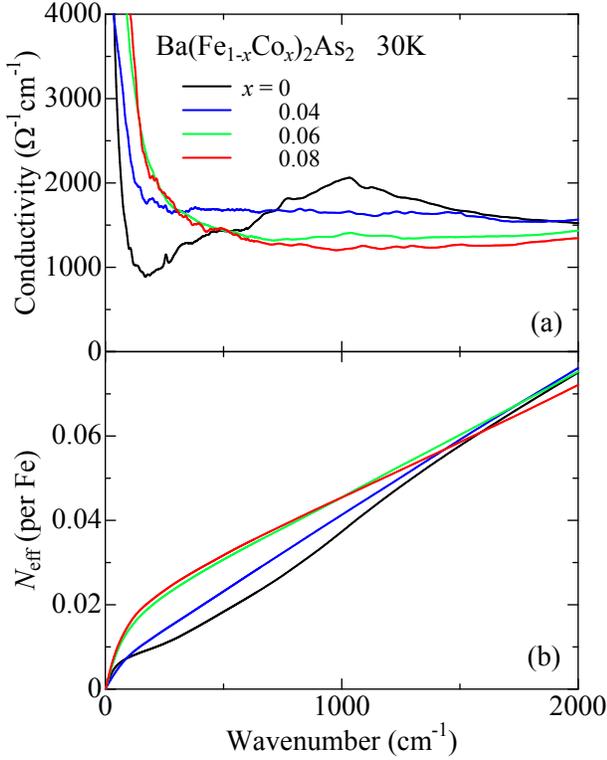}%
\caption{(Color online) Doping evolutions of (a) the optical conductivity spectrum and (b) the \Neff{} spectrum at $T$ = 30 K. \label{04}}
\end{figure}

A quite contrasting picture emerges when one sees the doping evolution of $\sigma_1 (\omega)$ starting from that in the magnetically ordered state of the undoped compound. In Fig.\ \ref{04}(a), we show $\sigma_1 (\omega)$ for the four compositions at 30 K. One finds a remarkable doping evolution of $\sigma_1 (\omega)$ in this case. It is quite similar to the temperature evolution of $\sigma_1 (\omega)$ for the parent compound, and is reminiscent of the high-\Tc{} cuprates where doping produces a dramatic transfer of the spectral weight over an eV energy range. \cite{Uchida1991} As is evidenced by \Neff($\omega$) at 30 K, the spectral weight transfer over a wide energy range is seen (Fig.\ \ref{04} (b)). A relatively large suppression of \Neff{} for $x$ = 0 in the low-energy region is due to the opening of the magnetic gap. \Neff{} in this region increases for $x$ = 0.04 as the gap is partially filled in. For $x$ = 0.06 and 0.08 \Neff{} is no longer suppressed, signaling a disappearance or a complete filling of the gap, in accord with the absence of the magnetic order at these doping levels (see Fig.\ \ref{01}). Note that all the \Neff{} curves merge at $\sim$ 1700 \cm{}, evidencing that the high-energy weight (up to 1700 \cm{}) in the magnetically ordered state transferred to lower-energy region, as the magnetic order is suppressed with doping. Although the energy scale is smaller by an order than that in the cuprates, a similar reconstruction of the electronic state should be taking place as a consequence of doping.

\section{Analysis of $\mathbf{\sigma_1 (\omega)}$ using a two-component model}

The optical conductivity spectrum above \TN{} is reminiscent of that for high-\Tc{} cuprates, and cannot be fit with a simple Drude term. In the case of cuprates, there are two ways of fitting the spectrum. One is to use so-called extended Drude formalism with a single Drude term with $\omega$-dependent scattering time $\tau(\omega)$. The other is to decompose the spectrum into two terms, a Drude term with $\omega$-independent $\tau$ and an ``incoherent'' (or ``mid-IR'') term which shows a broad peak in the mid-infrared region with a long high-energy tail. \cite{Thomas1988} In the present case it is possible to adopt the extended Drude, but $1/\tau (\omega)$ exhibits unrealistic feature at $\omega$ higher than $\sim$ 1000 \cm{}, since $\sigma_1 (\omega)$ is too flat or even increasing in the high-energy region. So, we fit the spectrum based on the multi-component analysis. Given that iron pnictides are multi-band/carrier systems, it is quite natural to use this analysis.

\subsection{Decomposition of $\sigma_1 (\omega)$ above \TN{}}

\begin{figure}
\includegraphics[width=80mm,clip]{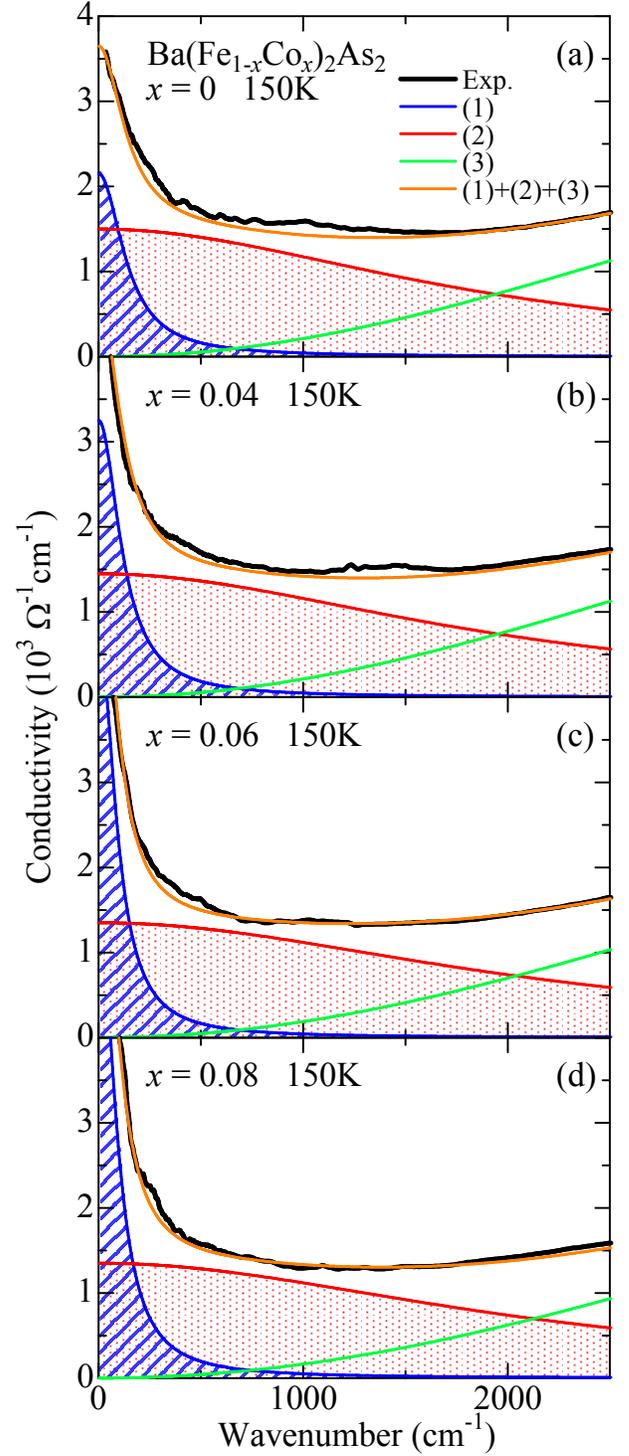}%
\caption{(Color online) Optical conductivity spectrum at $T$ = 150 K for (a) $x$ = 0, (b) 0.04, (c) 0.06, and (d) 0.08. Each spectrum is decomposed into a Drude term (blue hatched) and an ``incoherent'' term (red shaded). The green line indicates an interband component which is necessary to fit the data in the high energy region. \label{05}}
\end{figure}

\begin{figure}
\includegraphics[width=80mm,clip]{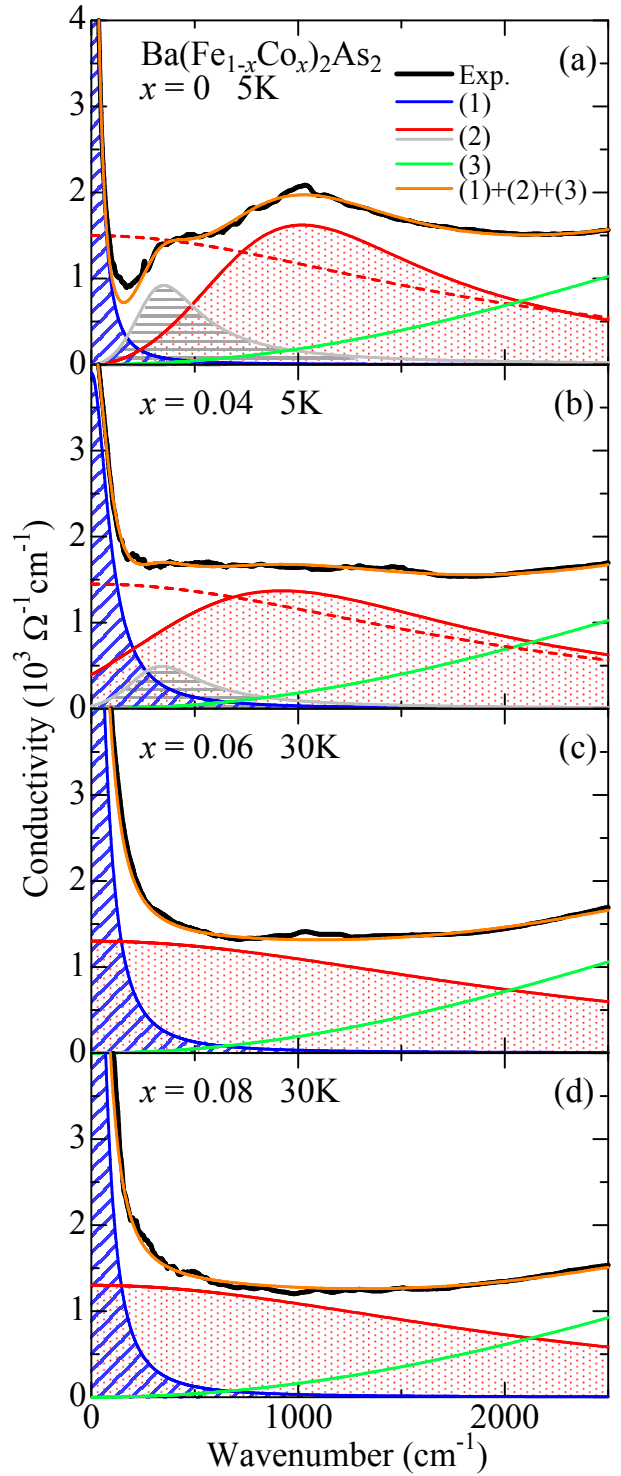}%
\caption{(Color online) Low-temperature optical conductivity spectrum and its decomposition for (a) $x$ = 0, (b) 0.04, (c) 0.06, and (d) 0.08. The data are at $T$ = 5 K for $x$ = 0 and 0.04, and at $T$ = 30 K for the others. A gap feature is seen in the ``incoherent'' term (red shaded) for $x$ = 0 and 0.04. An additional component (gray) is necessary to best fit the experimental result. For $x$ = 0.06 and 0.08, the ``incoherent'' term is almost unchanged as compared also with that at $T$ = 150 K. \label{06}}
\end{figure}

Figure \ref{05} shows the decomposition of the optical conductivity spectra at $T$ = 150 K for all the four compounds. In the low-energy region, $\sigma_1 (\omega)$ is dominated by two components. One component is a Drude term ($\sigma_{\mathrm{D}} (\omega)$) showing a relatively sharp peak at $\omega = 0$, and the other is an ``incoherent'' term ($\sigma_{\mathrm{in}} (\omega)$) with a long higher-energy tail. Since the peak at $\omega = 0$ is rather sharp, we first fit the spectrum in the lowest-energy region with a Drude term plus constant background (for doped compounds the $\omega = 0$ peak is so sharp that the isolation of the Drude term from $\sigma_1 (\omega)$ is unambiguously done). Next, we approximate the constant background by a single overdamped Drude term (we call this an ``incoherent'' term) to best fit the spectrum below 1000 \cm{} combined with the Drude component (shown by orange curves). There is a room to improve the fitting by adding one or more Drude-Lorentz terms with smaller weight, as will be the case with $\sigma_1 (\omega)$ at $T$ = 5 K (Fig.\ \ref{06}). In order to fit the spectrum in higher-energy region, we need to add a third term indicated by the green line which would correspond to an interband transition showing a peak around 7000 \cm{} (see Fig.\ \ref{03}(b)).

\begin{figure}
\includegraphics[width=80mm,clip]{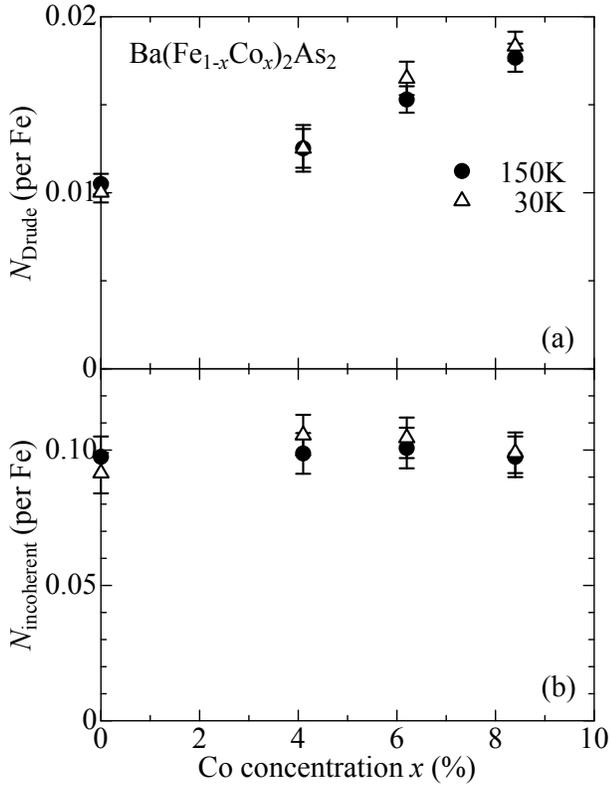}%
\caption{Spectral weights of (a) the Drude term and (b) the ``incoherent'' term plotted as a function of Co concentration at $T$ = 150 K (closed circles) and 30 K (open triangles). \label{07}}
\end{figure}

Different from the incoherent term in cuprates which exhibits a broad peak in the mid-IR region, the ``incoherent'' term in this case does not show a peak feature and contributes to the conductivity at $\omega$ = 0. We plot the spectral weight of the Drude and the ``incoherent'' components as a function of $x$ in Fig.\ \ref{07}. For the parent compound, the Drude weight is very small, about 10 \% of the ``incoherent'' weight. With increasing doping, the Drude component increases its weight, nearly doubled for $x$ = 0.08, while the ``incoherent'' component does not show appreciable doping dependence. The Drude weight even for $x$ = 0.08 is only a small fraction of the ``incoherent'' weight. Therefore, the weak doping dependence of \Neff($\omega$) at $T$ = 150 K shown in Fig.\ \ref{03}(c) is explained by the increase in the Drude component with doping, consistent with the picture that the Co substitution increases electron density responsible for the Drude term. This, in turn, gives evidence that the decomposition shown in Fig.\ \ref{05} is reasonable.

\subsection{Decomposition of $\sigma_1 (\omega)$ below \TN{}}

Then, we analyze the effect of the magnetic order on each conductivity component. For the undoped compound, the spectrum at 5 K well below \TN{} is decomposed in such a way as shown in Fig.\ \ref{06}(a). It turns out that the lowest-energy spectrum can be fitted using a narrowed Drude term with the same weight as that in $\sigma_1 (\omega)$ at $T$ = 150 K. Then, it follows that a gap should open in the ``incoherent'' component, and that $\sigma_{\mathrm{in}} (\omega)$ no longer contributes to the dc conductivity. At $T$ = 5 K, the presence of an additional small component (gray region), also showing a gap feature, becomes apparent, indicating that the ``incoherent'' term may originally consist of at least two components. As the Drude weight is only 10 \% of the total low-energy weight in $\sigma_1 (\omega)$ above \TN{}, nearly 90 \% of the low-energy weight contributing to the dc conductivity is lost by a gap opening below \TN{} (the same result is reported for CaFe$_2$As$_2$). \cite{Nakajima2009} This suggests that only a small piece of the multiple Fermi surfaces survives in the magnetically ordered state. Although no firm evidence is given from ARPES, the rapid narrowing of the Drude peak or rapid decrease in the carrier scattering rate below \TN{} (Fig.\ \ref{08} (a)) can be explained by a contraction of momentum space available for scattering due to disappearance of many portions of the Fermi-surface sheets.

For $x$ = 0.04, \TN{} goes down to 80 K. Below \TN{}, the gap feature in the spectrum is seen with the gap-energy scale nearly the same as that in the parent \BFA{}. Assuming again that the Drude weight does not change across \TN{}, we find that a gap opens in the ``incoherent'' component $\sigma_{\mathrm{in}} (\omega)$. However, it is not a complete gap, but a ``pseudogap.'' A small but finite conductivity remains at $\omega$ = 0, and $\sigma_{\mathrm{in}}$ contributes to the dc conductivity even below \TN{}.

With increase in doping level to $x$ = 0.06 and 0.08, the magnetic order is completely suppressed. For these compositions, because no gap feature is seen in $\sigma_1 (\omega)$ and the peak at $\omega = 0$ is so pronounced, the decomposition of $\sigma_1 (\omega)$ as shown in Figs.\ \ref{05} and \ref{06} is unique and robust. We can conclude that $\sigma_{\mathrm{in}} (\omega)$ is almost unchanged between $T$ = 150 and 30 K, and also between $x$ = 0.06 and 0.08. Note again that the ``incoherent'' spectral shape $\sigma_{\mathrm{in}} (\omega)$ at $T$ = 150 K as well as its weight both above and below \TN{} are nearly doping independent (Fig.\ \ref{07}). From the decomposed spectra shown in Figs.\ \ref{06}(a)-\ref{06}(d), $\sigma_{\mathrm{in}} (\omega)$ is found to be a good indicator for the presence or absence of the magnetic order. It should originate from the Fermi-surface sheets or portions of Fermi surfaces that undergo reconstruction in the magnetically ordered state.

\subsection{Carrier scattering rate and temperature dependence of dc resistivity}

\begin{figure}
\includegraphics[width=80mm,clip]{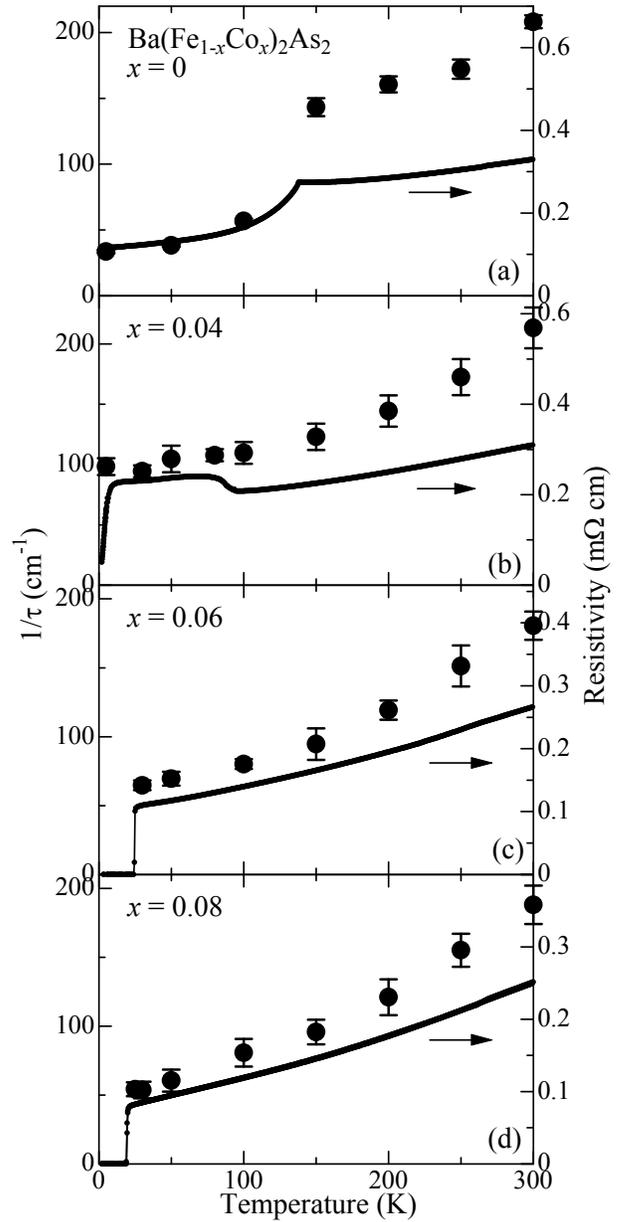}%
\caption{Width of the Drude term ($1/\tau$) is plotted as a function of temperature for (a) $x$ = 0, (b) 0.04, (c) 0.06, and (d) 0.08. For comparison, temperature dependence of resistivity is also shown (the right-hand scale). The scales of $1/\tau$ and resistivity are adjusted by taking into account the contribution of the ``incoherent'' component to the dc conductivity. \label{08}}
\end{figure}

The width of thus extracted Drude component, $\sigma_{\mathrm{D}} (\omega) = \sigma_{\mathrm{D}} (\omega=0) / (1 + \omega^2 \tau^2)$, corresponding to carrier scattering rate $1/\tau$ at various temperatures is plotted in Figs.\ \ref{08}(a)-\ref{08}(d) for all the four compounds. For comparison, $T$ dependence of resistivity is overlaid on each graph. The scales of Drude width and resistivity are adjusted by taking into account the contribution of $\sigma_{\mathrm{in}}$ to the conductivity at $\omega = 0$ using $\sigma_1 (\omega=0) = \sigma_{\mathrm{D}} (\omega=0) + \sigma_{\mathrm{in}} (\omega=0)$ and $\rho = 1/\sigma_1 (0)$. Both curves come close to each other in the low-$T$ region. This is because $\sigma_{\mathrm{D}}$ dominates, $\sigma_{\mathrm{D}} (\omega=0) \gg \sigma_{\mathrm{in}} (\omega=0)$, owing to the gap (or pseudogap) for $x$ = 0 and 0.04, and to relatively smaller contribution of the ``incoherent'' term for $x$ = 0.06 and 0.08. The large separation between the two in the high-$T$ region is due to a fairly large contribution from $\sigma_{\mathrm{in}}$ to the dc conductivity. For the parent compound, the $T$ dependence of resistivity is determined solely by the Drude term below \TN{}. Note that there exists a fairly large residual component in $1/\tau$ even in this undoped compound without dopant disorder. We speculate that the scattering source might be magnetic domain boundaries necessarily formed because of collinear (stripy) magnetic order. \cite{Ishida2009}

For $x$ = 0.04, different from the case with $x$ = 0, $1/\tau$ does not show a sharp drop at \TN{} nor strong $T$ dependence below \TN{} (Fig.\ \ref{08}(b)). Certainly, this is due to strong elastic scattering from Co impurities in the Fe plane, and also to incomplete reconstruction of Fermi surfaces, reflecting the weakened or disordered magnetic order. Then, it follows that the observed steplike increase in resistivity at \TN{} is associated with the decrease in the contribution from $\sigma_{\mathrm{in}}$ due to the pseudogap. For the superconducting compositions (Figs.\ \ref{08}(c) and \ref{08}(d)), $T$ dependence of the scattering rate is also extrapolated to a considerably large value at $T$ = 0 K. In these cases, the carrier scattering centers should be dopant Co atoms randomly substituted for Fe atoms. The residual component is largest for $x$ = 0.04 in which scattering from magnetic-domain boundaries would add to scattering from Co impurities.

As regards an inelastic scattering component in $1/\tau$ of highly doped compounds ($x$ = 0.06 and 0.08), it shows power-law $T$ dependence ($\sim T^n$) above \Tc{} with an exponent $n$ seemingly larger than 1 because of its concave curvature (it is also possible to see the data as a crossover from constant at low temperatures to $T$-linear at high temperatures). An obvious difference from the $T$ dependence of resistivity, which seems to be dominated by a $T$-linear term, arises from the contribution from $\sigma_{\mathrm{in}} (\omega)$. According to the present analysis, $\sigma_{\mathrm{in}} (\omega)$ and hence $\sigma_{\mathrm{in}} (0)$ are nearly $T$ independent, so $\sigma_{\mathrm{in}} (0)$ increasingly contributes to resistivity as temperature is raised, making $T$ dependence of $\rho$ weaker than that of $1/\tau$ extracted from the Drude component. Down-bending of $\rho(T)$ frequently observed for $Ln$FeAsO family ($Ln$ = lanthanide elements) \cite{Miyazawa2009} and K-substituted \BFA{} \cite{Chen2008} might be explained by dominant contribution of $\sigma_{\mathrm{in}} (0)$ in the higher-temperature region.

Wu \textit{et al.}\ \cite{Wu2009b} performed similar two-component analysis of the optical conductivity spectra of the same compounds with $x$ = 0 (also EuFe$_2$As$_2$) and 0.08. When the present data are compared with theirs, one realizes quantitative differences between the two. A critical difference arises in that the measured reflectivity in the lowest-energy region in Ref.~\onlinecite{Wu2009b} is too high as compared with the present result. Hence, the low-energy conductivity (and consequently the Drude spectral weight in the spectrum) at $T$ = 150 (210) K for \BFA{} (EuFe$_2$As$_2$) is much bigger than the present one, which leads Wu \textit{et al.}\ to suppose that a gap below \TN{} would open predominantly in $\sigma_{\mathrm{D}} (\omega)$, which reduces the Drude weight and transfers it to higher energies. The bigger Drude weight of their analysis results from a broader peak at $\omega=0$ for EuFe$_2$As$_2$ measured at relatively high temperature $T$ = 210 K (\TN{} = 189 K in this compound). \cite{Wu2009} In fact, it is possible to decompose the present spectrum at $T$ = 150 K for $x$ = 0 into the Drude component with bigger weight, and hence smaller weight for the ``incoherent'' component, although the fitting to the experimental data is a little bit worse. However, in view of the systematic variations in the Drude component with doping and temperature as well as remarkable consistency with the dc resistivity, we conclude that the present decomposition is more reliable and reasonable.

\subsection{$\sigma_1 (\omega)$ in the SC state}

\begin{figure}
\includegraphics[width=80mm,clip]{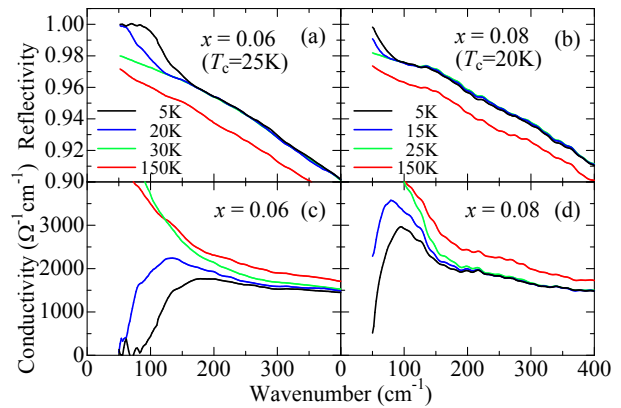}%
\caption{(Color online) Reflectivity spectrum and optical conductivity spectrum of \BFCA{} for (a) (c) $x$ =0.06 and (b) (d) 0.08 above and below the superconducting transition temperature \Tc{}. \label{09}}
\end{figure}

Finally, we focus on the optical properties in the SC state. Reflectivity spectra for $x$ = 0.06 and 0.08 are presented in Figs.\ \ref{09}(a) and \ref{09}(b), respectively. The evolution of the spectrum with temperature is similar to that of K-doped \BFA{}. \cite{Li2008} Below \Tc{}, reflectivity of the compound with $x$ = 0.06 (0.08) exhibits enhancement to almost unity due to SC gap opening below 80 (50) \cm{}. As evidenced by the optical conductivity spectra shown in Figs.\ \ref{09}(c) and \ref{09}(d), conductivity in this energy region is radically suppressed, suggesting that Co-doped \BFA{} is a full gap superconductor in agreement with ARPES \cite{Terashima2009} and also with theoretical proposals. \cite{Mazin2008, Kuroki2008} Given that the low-energy spectrum contains some ambiguity due to experimental uncertainty and limitation in the energy region covered in the experiment, we cannot rule out a possibility that a second weak gap feature is hidden above the conductivity edge at $\sim$ 80 \cm{} \cite{Kim2009, Heumen2009} or that finite conductivity remains in the region below 80 \cm{} which may show a gap feature indicative of a second smaller gap. \cite{Wu2009b, Gorshunov2009} However, the steep drop of conductivity below 120 \cm{} toward 80 \cm{} is a robust result. So, if we take this as an $s$-wave gap edge, the gap magnitude would be 2$\Delta \sim$ 80 \cm{} at $T$ = 5 K for $x$ = 0.06 corresponding to $2\Delta/ k_{\mathrm{B}} T_{\mathrm{c}} \sim 4.6$ (probably, $2\Delta$ for $x$ = 0.08 is 50 \cm{} or lower). This value is very close to the value estimated by ARPES \cite{Terashima2009} and by the point contact measurement. \cite{Samuely2009} In the case of K-doped \BFA{} with \Tc{} = 37 K, a similarly defined gap $2\Delta$ is about 150 \cm{} ($2\Delta/ k_{\mathrm{B}} T_{\mathrm{c}} \sim 5.8$). \cite{Li2008} We see that $2\Delta$ is larger for higher \Tc{} in doped \BFA{}, but we are not sure if $2\Delta$ exactly scales with \Tc{}.

From the missing area of the conductivity spectrum in the SC state, we estimate the superfluid density or condensed weight at $\omega$ = 0;
\begin{equation}
\rho_s = 8 \int^{\omega_c}_{0^+} \mathrm{d}\omega [ \sigma_1 (\omega, T \simeq T_{\mathrm{c}}) - \sigma_1 (\omega, T = 5 \mathrm{K})].
\end{equation}
The London penetration depth $\lambda_{\mathrm{L}}$ is related to the superfluid density by $\rho_s = c^2 / \lambda_{\mathrm{L}} ^2$. Calculating the missing area (setting $\omega_c \sim$ 500 \cm{}) from our results, $\lambda_{\mathrm{L}}$ is estimated to be 2770 $\pm$ 250 (3150 $\pm$ 300) \AA{} for $x$ = 0.06 (0.08). These values are in good agreement with the value provided by the muon spin relaxation \cite{Williams2009} and the tunnel diode resonator technique. \cite{Gordon2009} Therefore, the Ferrell-Glover-Tinkham sum rule appears to hold in the present SC system.

\begin{figure}
\includegraphics[width=80mm,clip]{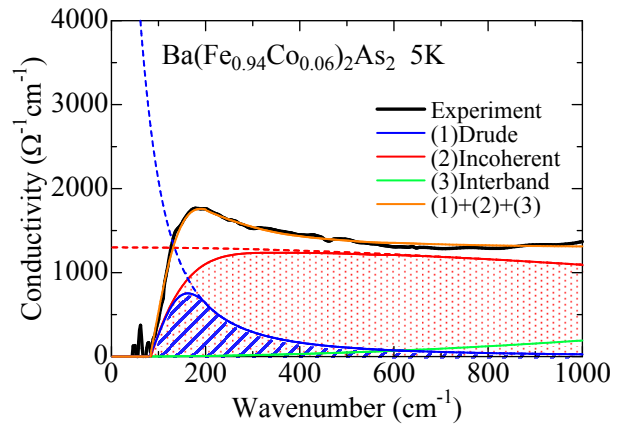}%
\caption{(Color online) Decomposition of the conductivity spectrum of \BFCAb{} at $T$ = 5 K. The experimental spectrum is best reproduced assuming a single SC gap of 80 \cm{}. The dashed lines indicate each component just above \Tc{}. \label{10}}
\end{figure}

A typical result of two-component analysis of the spectrum in the SC state is illustrated in Fig.\ \ref{10}. The blue and red dashed lines reproduce $\sigma_{\mathrm{D}} (\omega)$ and $\sigma_{\mathrm{in}} (\omega)$ at $T$ = 30 K just above \Tc{}. The spectrum in the SC state at $T$ = 5 K cannot be reproduced unless the SC gap affects both components. A decomposition into the blue solid curve (hatched) and the red solid curve (shaded) best fits the experimental spectrum at $T$ = 5 K. Incorporation of multiple SC gaps is not necessary as long as the level of fitting is the one shown by the orange curve in Fig.\ \ref{10}. The result suggests that a full gap opens in every piece of Fermi-surface sheets in the SC state.

\subsection{Possible origin of ``incoherent'' term}

The Drude component certainly has its origin in the electron Fermi surfaces as its weight increases with Co content. In the magnetically ordered state the gap opens only in the ``incoherent'' component, and this gap just follows the evolution of the magnetic order with doping. In this respect, the ``incoherent'' component would arise from some Fermi-surface sheets or their segments which fulfill the nesting condition for an SDW instability. Alternatively, the selective gap opening and its intimate connection to the magnetic order may imply that the relevant Fermi surfaces are dominated by particular Fe 3$d$ orbitals which are involved in the formation of the magnetic order. It is speculated that the magnetic order is accompanied with an orbital ordering, \cite{Shimojima2009} since a structural transition always occurs at or before magnetic ordering in many of the iron arsenides. One should note that the ``incoherent'' component persists into the SC regime without showing gap feature but with almost unchanged spectral weight. This suggests that persistent spin/orbital fluctuations strongly scatter the charge carriers in the normal state of the SC compounds. A similar doping evolution of magnetism and electronic structure in this system is also discussed by Canfield \textit{et al.}\ \cite{Canfield2009} based on the transport study.

\section{Summary}

A systematic investigation of the optical spectrum is performed on \BFCA{}. The optical conductivity spectrum in the non-magnetic state of the undoped compound changes weakly with doping. By contrast, the spectrum at temperatures below \TN{} exhibits a dramatic doping evolution over an energy range up to 2000 \cm{}. The high-energy spectral weight associated with an SDW/magnetic gap is transferred into the low-energy region, reminiscent of the doping evolution of the optical spectrum for high-\Tc{} cuprates and indicative of a reconstruction of the electronic state on large-energy scale.

The low-energy spectrum can be decomposed into two components, a Drude and an ``incoherent'' term, reflecting the presence of two or more distinct charge carriers in iron-arsenic systems. Temperature and doping evolutions of the Drude component combined with those of the ``incoherent'' component are shown to well explain the temperature dependence of dc resistivity at various doping levels. The Drude component increases its weight with doping obviously originating from electron Fermi surfaces, whereas the weight of the ``incoherent'' component does not appreciably change. It is demonstrated that a gap in the magnetic state opens selectively in the ``incoherent'' component, and that the gap follows the evolution of the magnetic order with doping. The ``incoherent'' component is thus considered to arise from Fermi surfaces that are relevant to the magnetic/orbital order, the fluctuations of which may persist well into the SC doping regime and cause strong carrier scattering. In the SC state, both Drude and ``incoherent'' components contribute to the SC condensation with an $s$-wave like gap opening in both components.

\textit{Note added}: During the completion of the manuscript a preprint has been up in the arXiv which reports the anisotropic charge transport in detwinned single crystals of Ba(Fe$_{1-x}$Co$_{x}$)$_2$As$_2$. \cite{Chu2010} The anisotropy in the $ab$-plane resistivity below $T \sim$ \TN{} is most pronounced for lightly doped compounds with $x$ in the range between 0.025 and 0.045. The resistivity in one direction ($b$ axis along which spins
allign ferromagnetically) increases with lowering temperature and tends to saturate at lower temperatures. On the other hand, the resistivity in the other direction ($a$ axis along which spins allign antiferromagnetically) continues to decrease from above \TN{}. In the light of the present result and analysis for $x$ = 0.04, this contrasting behavior is explained by a gap/pseudogap opening in $\sigma_{\mathrm{in}} (\omega)$ selectively for polarization parallel to the $b$ axis.


\begin{acknowledgments}

We would like to thank M.~Mamiya for ICP analysis. This work was supported by Transformative Research-Project on Iron Pnictides (TRIP) from the Japan Science and Technology Agency, and by the Global Centers of Excellence Program and the Japan-China-Korea A3 Foresight Program from Japan Society for the Promotion of Science, and a Grant-in-Aid of Scientific Research on Innovative Areas ``Heavy Electrons" (No.\ 20102005) from the Ministry of Education, Culture, Sports, Science, and Technology in Japan.

\end{acknowledgments}

\bibliographystyle{apsrev4-1}
\bibliography{Co-Ba122_ver12}

\begin{thebibliography}{10}%
\makeatletter
\providecommand \@ifxundefined [1]{%
 \ifx #1\undefined \expandafter \@firstoftwo
 \else \expandafter \@secondoftwo
\fi
}%
\providecommand \@ifnum [1]{%
 \ifnum #1\expandafter \@firstoftwo
 \else \expandafter \@secondoftwo
\fi
}%
\providecommand \enquote [1]{``#1''}%
\providecommand \bibnamefont  [1]{#1}%
\providecommand \bibfnamefont [1]{#1}%
\providecommand \citenamefont [1]{#1}%
\providecommand\href[0]{\@sanitize\@href}%
\providecommand\@href[1]{\endgroup\@@startlink{#1}\endgroup\@@href}%
\providecommand\@@href[1]{#1\@@endlink}%
\providecommand \@sanitize [0]{\begingroup\catcode`\&12\catcode`\#12\relax}%
\@ifxundefined \pdfoutput {\@firstoftwo}{%
 \@ifnum{\z@=\pdfoutput}{\@firstoftwo}{\@secondoftwo}%
}{%
 \providecommand\@@startlink[1]{\leavevmode\special{html:<a href="#1">}}%
 \providecommand\@@endlink[0]{\special{html:</a>}}%
}{%
 \providecommand\@@startlink[1]{%
  \leavevmode
  \pdfstartlink
   attr{/Border[0 0 1 ]/H/I/C[0 1 1]}%
   user{/Subtype/Link/A<</Type/Action/S/URI/URI(#1)>>}%
  \relax
 }%
 \providecommand\@@endlink[0]{\pdfendlink}%
}%
\providecommand \url  [0]{\begingroup\@sanitize \@url }%
\providecommand \@url [1]{\endgroup\@href {#1}{\urlprefix}}%
\providecommand \urlprefix [0]{URL }%
\providecommand \Eprint[0]{\href }%
\@ifxundefined \urlstyle {%
  \providecommand \doi [1]{doi:\discretionary{}{}{}#1}%
}{%
  \providecommand \doi [0]{doi:\discretionary{}{}{}\begingroup
  \urlstyle{rm}\Url }%
}%
\providecommand \doibase [0]{http://dx.doi.org/}%
\providecommand \Doi[1]{\href{\doibase#1}}%
\providecommand \bibAnnote [3]{%
  \BibitemShut{#1}%
  \begin{quotation}\noindent
    \textsc{Key:}\ #2\\\textsc{Annotation:}\ #3%
  \end{quotation}%
}%
\providecommand \bibAnnoteFile [2]{%
  \IfFileExists{#2}{\bibAnnote {#1} {#2} {\input{#2}}}{}%
}%
\providecommand \typeout [0]{\immediate \write \m@ne }%
\providecommand \selectlanguage [0]{\@gobble}%
\providecommand \bibinfo [0]{\@secondoftwo}%
\providecommand \bibfield [0]{\@secondoftwo}%
\providecommand \translation [1]{[#1]}%
\providecommand \BibitemOpen[0]{}%
\providecommand \bibitemStop [0]{}%
\providecommand \bibitemNoStop [0]{.\EOS\space}%
\providecommand \EOS [0]{\spacefactor3000\relax}%
\providecommand \BibitemShut [1]{\csname bibitem#1\endcsname}%
\bibitem{Zhao2008}%
  \BibitemOpen
  \bibfield{author}{%
  \bibinfo {author} {\bibfnamefont{J.}~\bibnamefont{Zhao}}, \bibinfo {author}
  {\bibfnamefont{Q.}~\bibnamefont{Huang}}, \bibinfo {author}
  {\bibfnamefont{C.}~\bibnamefont{de~la Cruz}}, \bibinfo {author}
  {\bibfnamefont{S.}~\bibnamefont{Li}}, \bibinfo {author}
  {\bibfnamefont{J.~W.}\ \bibnamefont{Lynn}}, \bibinfo {author}
  {\bibfnamefont{Y.}~\bibnamefont{Chen}}, \bibinfo {author}
  {\bibfnamefont{M.~A.}\ \bibnamefont{Green}}, \bibinfo {author}
  {\bibfnamefont{G.~F.}\ \bibnamefont{Chen}}, \bibinfo {author}
  {\bibfnamefont{G.}~\bibnamefont{Li}}, \bibinfo {author}
  {\bibfnamefont{Z.}~\bibnamefont{Li}}, \bibinfo {author}
  {\bibfnamefont{J.~L.}\ \bibnamefont{Luo}}, \bibinfo {author}
  {\bibfnamefont{N.~L.}\ \bibnamefont{Wang}},\ and\ \bibinfo {author}
  {\bibfnamefont{P.}~\bibnamefont{Dai}},\ }%
  \bibfield{journal}{%
  \bibinfo {journal} {Nature Mater.}\ }%
  \textbf{\bibinfo {volume} {7}},\ \bibinfo {pages} {953} (\bibinfo {year}
  {2008})%
  \bibAnnoteFile{NoStop}{Zhao2008}%
\bibitem{Chen2008}%
  \BibitemOpen
  \bibfield{author}{%
  \bibinfo {author} {\bibfnamefont{H.}~\bibnamefont{Chen}}, \bibinfo {author}
  {\bibfnamefont{Y.}~\bibnamefont{Ren}}, \bibinfo {author}
  {\bibfnamefont{Y.}~\bibnamefont{Qiu}}, \bibinfo {author}
  {\bibfnamefont{W.}~\bibnamefont{Bao}}, \bibinfo {author}
  {\bibfnamefont{R.~H.}\ \bibnamefont{Liu}}, \bibinfo {author}
  {\bibfnamefont{G.}~\bibnamefont{Wu}}, \bibinfo {author}
  {\bibfnamefont{T.}~\bibnamefont{Wu}}, \bibinfo {author}
  {\bibfnamefont{Y.~L.}\ \bibnamefont{Xie}}, \bibinfo {author}
  {\bibfnamefont{X.~F.}\ \bibnamefont{Wang}}, \bibinfo {author}
  {\bibfnamefont{Q.}~\bibnamefont{Huang}},\ and\ \bibinfo {author}
  {\bibfnamefont{X.~H.}\ \bibnamefont{Chen}},\ }%
  \bibfield{journal}{%
  \bibinfo {journal} {Europhys. Lett.}\ }%
  \textbf{\bibinfo {volume} {85}},\ \bibinfo {pages} {17006} (\bibinfo {year}
  {2009})%
  \bibAnnoteFile{NoStop}{Chen2008}%
\bibitem{Nandi2009}%
  \BibitemOpen
  \bibfield{author}{%
  \bibinfo {author} {\bibfnamefont{S.}~\bibnamefont{Nandi}}, \bibinfo {author}
  {\bibfnamefont{M.~G.}\ \bibnamefont{Kim}}, \bibinfo {author}
  {\bibfnamefont{A.}~\bibnamefont{Kreyssig}}, \bibinfo {author}
  {\bibfnamefont{R.~M.}\ \bibnamefont{Fernandes}}, \bibinfo {author}
  {\bibfnamefont{D.~K.}\ \bibnamefont{Pratt}}, \bibinfo {author}
  {\bibfnamefont{A.}~\bibnamefont{Thaler}}, \bibinfo {author}
  {\bibfnamefont{N.}~\bibnamefont{Ni}}, \bibinfo {author}
  {\bibfnamefont{S.~L.}\ \bibnamefont{Bud'ko}}, \bibinfo {author}
  {\bibfnamefont{P.~C.}\ \bibnamefont{Canfield}}, \bibinfo {author}
  {\bibfnamefont{J.}~\bibnamefont{Schmalian}}, \bibinfo {author}
  {\bibfnamefont{R.~J.}\ \bibnamefont{McQueeney}},\ and\ \bibinfo {author}
  {\bibfnamefont{A.~I.}\ \bibnamefont{Goldman}},\ }%
  \bibfield{journal}{%
  \bibinfo {journal} {Phys. Rev. Lett.}\ }%
  \textbf{\bibinfo {volume} {104}},\ \bibinfo {pages} {057006} (\bibinfo {year}
  {2010})%
  \bibAnnoteFile{NoStop}{Nandi2009}%
\bibitem{Mazin2008}%
  \BibitemOpen
  \bibfield{author}{%
  \bibinfo {author} {\bibfnamefont{I.~I.}\ \bibnamefont{Mazin}}, \bibinfo
  {author} {\bibfnamefont{D.~J.}\ \bibnamefont{Singh}}, \bibinfo {author}
  {\bibfnamefont{M.~D.}\ \bibnamefont{Johannes}},\ and\ \bibinfo {author}
  {\bibfnamefont{M.~H.}\ \bibnamefont{Du}},\ }%
  \bibfield{journal}{%
  \bibinfo {journal} {Phys. Rev. Lett.}\ }%
  \textbf{\bibinfo {volume} {101}},\ \bibinfo {pages} {057003} (\bibinfo {year}
  {2008})%
  \bibAnnoteFile{NoStop}{Mazin2008}%
\bibitem{Kuroki2008}%
  \BibitemOpen
  \bibfield{author}{%
  \bibinfo {author} {\bibfnamefont{K.}~\bibnamefont{Kuroki}}, \bibinfo {author}
  {\bibfnamefont{S.}~\bibnamefont{Onari}}, \bibinfo {author}
  {\bibfnamefont{R.}~\bibnamefont{Arita}}, \bibinfo {author}
  {\bibfnamefont{H.}~\bibnamefont{Usui}}, \bibinfo {author}
  {\bibfnamefont{Y.}~\bibnamefont{Tanaka}}, \bibinfo {author}
  {\bibfnamefont{H.}~\bibnamefont{Kontani}},\ and\ \bibinfo {author}
  {\bibfnamefont{H.}~\bibnamefont{Aoki}},\ }%
  \bibfield{journal}{%
  \bibinfo {journal} {Phys. Rev. Lett.}\ }%
  \textbf{\bibinfo {volume} {101}},\ \bibinfo {pages} {087004} (\bibinfo {year}
  {2008})%
  \bibAnnoteFile{NoStop}{Kuroki2008}%
\bibitem{Hsieh2008}%
  \BibitemOpen
  \bibfield{author}{%
  \bibinfo {author} {\bibfnamefont{D.}~\bibnamefont{Hsieh}}, \bibinfo {author}
  {\bibfnamefont{Y.}~\bibnamefont{Xia}}, \bibinfo {author}
  {\bibfnamefont{L.}~\bibnamefont{Wray}}, \bibinfo {author}
  {\bibfnamefont{D.}~\bibnamefont{Qian}}, \bibinfo {author}
  {\bibfnamefont{K.~K.}\ \bibnamefont{Gomes}}, \bibinfo {author}
  {\bibfnamefont{A.}~\bibnamefont{Yazdani}}, \bibinfo {author}
  {\bibfnamefont{G.~F.}\ \bibnamefont{Chen}}, \bibinfo {author}
  {\bibfnamefont{J.~L.}\ \bibnamefont{Luo}}, \bibinfo {author}
  {\bibfnamefont{N.~L.}\ \bibnamefont{Wang}},\ and\ \bibinfo {author}
  {\bibfnamefont{M.~Z.}\ \bibnamefont{Hasan}},\ }%
  \bibinfo {note} {arXiv:0812.2289 (unpublished)}%
  \bibAnnoteFile{NoStop}{Hsieh2008}%
\bibitem{Yi2009}%
  \BibitemOpen
  \bibfield{author}{%
  \bibinfo {author} {\bibfnamefont{M.}~\bibnamefont{Yi}}, \bibinfo {author}
  {\bibfnamefont{D.~H.}\ \bibnamefont{Lu}}, \bibinfo {author}
  {\bibfnamefont{J.~G.}\ \bibnamefont{Analytis}}, \bibinfo {author}
  {\bibfnamefont{J.-H.}\ \bibnamefont{Chu}}, \bibinfo {author}
  {\bibfnamefont{S.-K.}\ \bibnamefont{Mo}}, \bibinfo {author}
  {\bibfnamefont{R.-H.}\ \bibnamefont{He}}, \bibinfo {author}
  {\bibfnamefont{M.}~\bibnamefont{Hashimoto}}, \bibinfo {author}
  {\bibfnamefont{R.~G.}\ \bibnamefont{Moore}}, \bibinfo {author}
  {\bibfnamefont{I.~I.}\ \bibnamefont{Mazin}}, \bibinfo {author}
  {\bibfnamefont{D.~J.}\ \bibnamefont{Singh}}, \bibinfo {author}
  {\bibfnamefont{Z.}~\bibnamefont{Hussain}}, \bibinfo {author}
  {\bibfnamefont{I.~R.}\ \bibnamefont{Fisher}},\ and\ \bibinfo {author}
  {\bibfnamefont{Z.-X.}\ \bibnamefont{Shen}},\ }%
  \bibfield{journal}{%
  \bibinfo {journal} {Phys. Rev. B}\ }%
  \textbf{\bibinfo {volume} {80}},\ \bibinfo {pages} {174510} (\bibinfo {year}
  {2009})%
  \bibAnnoteFile{NoStop}{Yi2009}%
\bibitem{Shimojima2009}%
  \BibitemOpen
  \bibfield{author}{%
  \bibinfo {author} {\bibfnamefont{T.}~\bibnamefont{Shimojima}}, \bibinfo
  {author} {\bibfnamefont{K.}~\bibnamefont{Ishizaka}}, \bibinfo {author}
  {\bibfnamefont{Y.}~\bibnamefont{Ishida}}, \bibinfo {author}
  {\bibfnamefont{N.}~\bibnamefont{Katayama}}, \bibinfo {author}
  {\bibfnamefont{K.}~\bibnamefont{Ohgushi}}, \bibinfo {author}
  {\bibfnamefont{T.}~\bibnamefont{Kiss}}, \bibinfo {author}
  {\bibfnamefont{M.}~\bibnamefont{Okawa}}, \bibinfo {author}
  {\bibfnamefont{T.}~\bibnamefont{Togashi}}, \bibinfo {author}
  {\bibfnamefont{X.-Y.}\ \bibnamefont{Wang}}, \bibinfo {author}
  {\bibfnamefont{C.-T.}\ \bibnamefont{Chen}}, \bibinfo {author}
  {\bibfnamefont{S.}~\bibnamefont{Watanabe}}, \bibinfo {author}
  {\bibfnamefont{R.}~\bibnamefont{Kadota}}, \bibinfo {author}
  {\bibfnamefont{T.}~\bibnamefont{Oguchi}}, \bibinfo {author}
  {\bibfnamefont{A.}~\bibnamefont{Chainani}},\ and\ \bibinfo {author}
  {\bibfnamefont{S.}~\bibnamefont{Shin}},\ }%
  \bibfield{journal}{%
  \bibinfo {journal} {Phys. Rev. Lett.}\ }%
  \textbf{\bibinfo {volume} {104}},\ \bibinfo {pages} {057002} (\bibinfo {year}
  {2010})%
  \bibAnnoteFile{NoStop}{Shimojima2009}%
\bibitem{Kotegawa2009}%
  \BibitemOpen
  \bibfield{author}{%
  \bibinfo {author} {\bibfnamefont{H.}~\bibnamefont{Kotegawa}}, \bibinfo
  {author} {\bibfnamefont{H.}~\bibnamefont{Sugawara}},\ and\ \bibinfo {author}
  {\bibfnamefont{H.}~\bibnamefont{Tou}},\ }%
  \bibfield{journal}{%
  \bibinfo {journal} {J. Phys. Soc. Jpn.}\ }%
  \textbf{\bibinfo {volume} {78}},\ \bibinfo {pages} {013709} (\bibinfo {year}
  {2009})%
  \bibAnnoteFile{NoStop}{Kotegawa2009}%
\bibitem{Kasahara2009}%
  \BibitemOpen
  \bibfield{author}{%
  \bibinfo {author} {\bibfnamefont{S.}~\bibnamefont{Kasahara}}, \bibinfo
  {author} {\bibfnamefont{T.}~\bibnamefont{Shibauchi}}, \bibinfo {author}
  {\bibfnamefont{K.}~\bibnamefont{Hashimoto}}, \bibinfo {author}
  {\bibfnamefont{K.}~\bibnamefont{Ikeda}}, \bibinfo {author}
  {\bibfnamefont{S.}~\bibnamefont{Tonegawa}}, \bibinfo {author}
  {\bibfnamefont{H.}~\bibnamefont{Ikeda}}, \bibinfo {author}
  {\bibfnamefont{H.}~\bibnamefont{Takeya}}, \bibinfo {author}
  {\bibfnamefont{K.}~\bibnamefont{Hirata}}, \bibinfo {author}
  {\bibfnamefont{T.}~\bibnamefont{Terashima}},\ and\ \bibinfo {author}
  {\bibfnamefont{Y.}~\bibnamefont{Matsuda}},\ }%
  \bibinfo {note} {arXiv:0905.4427 (unpublished)}%
  \bibAnnoteFile{NoStop}{Kasahara2009}%
\bibitem{Hu2008}%
  \BibitemOpen
  \bibfield{author}{%
  \bibinfo {author} {\bibfnamefont{W.~Z.}\ \bibnamefont{Hu}}, \bibinfo {author}
  {\bibfnamefont{J.}~\bibnamefont{Dong}}, \bibinfo {author}
  {\bibfnamefont{G.}~\bibnamefont{Li}}, \bibinfo {author}
  {\bibfnamefont{Z.}~\bibnamefont{Li}}, \bibinfo {author}
  {\bibfnamefont{P.}~\bibnamefont{Zheng}}, \bibinfo {author}
  {\bibfnamefont{G.~F.}\ \bibnamefont{Chen}}, \bibinfo {author}
  {\bibfnamefont{J.~L.}\ \bibnamefont{Luo}},\ and\ \bibinfo {author}
  {\bibfnamefont{N.~L.}\ \bibnamefont{Wang}},\ }%
  \bibfield{journal}{%
  \bibinfo {journal} {Phys. Rev. Lett.}\ }%
  \textbf{\bibinfo {volume} {101}},\ \bibinfo {pages} {257005} (\bibinfo {year}
  {2008})%
  \bibAnnoteFile{NoStop}{Hu2008}%
\bibitem{Wu2009}%
  \BibitemOpen
  \bibfield{author}{%
  \bibinfo {author} {\bibfnamefont{D.}~\bibnamefont{Wu}}, \bibinfo {author}
  {\bibfnamefont{N.}~\bibnamefont{Bari\v{s}i\'{c}}}, \bibinfo {author}
  {\bibfnamefont{N.}~\bibnamefont{Drichko}}, \bibinfo {author}
  {\bibfnamefont{S.}~\bibnamefont{Kaiser}}, \bibinfo {author}
  {\bibfnamefont{A.}~\bibnamefont{Faridian}}, \bibinfo {author}
  {\bibfnamefont{M.}~\bibnamefont{Dressel}}, \bibinfo {author}
  {\bibfnamefont{S.}~\bibnamefont{Jiang}}, \bibinfo {author}
  {\bibfnamefont{Z.}~\bibnamefont{Ren}}, \bibinfo {author}
  {\bibfnamefont{L.~J.}\ \bibnamefont{Li}}, \bibinfo {author}
  {\bibfnamefont{G.~H.}\ \bibnamefont{Cao}}, \bibinfo {author}
  {\bibfnamefont{Z.~A.}\ \bibnamefont{Xu}}, \bibinfo {author}
  {\bibfnamefont{H.~S.}\ \bibnamefont{Jeevan}},\ and\ \bibinfo {author}
  {\bibfnamefont{P.}~\bibnamefont{Gegenwart}},\ }%
  \bibfield{journal}{%
  \bibinfo {journal} {Phys. Rev. B}\ }%
  \textbf{\bibinfo {volume} {79}},\ \bibinfo {pages} {155103} (\bibinfo {year}
  {2009})%
  \bibAnnoteFile{NoStop}{Wu2009}%
\bibitem{Li2008}%
  \BibitemOpen
  \bibfield{author}{%
  \bibinfo {author} {\bibfnamefont{G.}~\bibnamefont{Li}}, \bibinfo {author}
  {\bibfnamefont{W.~Z.}\ \bibnamefont{Hu}}, \bibinfo {author}
  {\bibfnamefont{J.}~\bibnamefont{Dong}}, \bibinfo {author}
  {\bibfnamefont{Z.}~\bibnamefont{Li}}, \bibinfo {author}
  {\bibfnamefont{P.}~\bibnamefont{Zheng}}, \bibinfo {author}
  {\bibfnamefont{G.~F.}\ \bibnamefont{Chen}}, \bibinfo {author}
  {\bibfnamefont{J.~L.}\ \bibnamefont{Luo}},\ and\ \bibinfo {author}
  {\bibfnamefont{N.~L.}\ \bibnamefont{Wang}},\ }%
  \bibfield{journal}{%
  \bibinfo {journal} {Phys. Rev. Lett.}\ }%
  \textbf{\bibinfo {volume} {101}},\ \bibinfo {pages} {107004} (\bibinfo {year}
  {2008})%
  \bibAnnoteFile{NoStop}{Li2008}%
\bibitem{Ni2008}%
  \BibitemOpen
  \bibfield{author}{%
  \bibinfo {author} {\bibfnamefont{N.}~\bibnamefont{Ni}}, \bibinfo {author}
  {\bibfnamefont{M.~E.}\ \bibnamefont{Tillman}}, \bibinfo {author}
  {\bibfnamefont{J.-Q.}\ \bibnamefont{Yan}}, \bibinfo {author}
  {\bibfnamefont{A.}~\bibnamefont{Kracher}}, \bibinfo {author}
  {\bibfnamefont{S.~T.}\ \bibnamefont{Hannahs}}, \bibinfo {author}
  {\bibfnamefont{S.~L.}\ \bibnamefont{Bud'ko}},\ and\ \bibinfo {author}
  {\bibfnamefont{P.~C.}\ \bibnamefont{Canfield}},\ }%
  \bibfield{journal}{%
  \bibinfo {journal} {Phys. Rev. B}\ }%
  \textbf{\bibinfo {volume} {78}},\ \bibinfo {pages} {214515} (\bibinfo {year}
  {2008})%
  \bibAnnoteFile{NoStop}{Ni2008}%
\bibitem{Chu2009}%
  \BibitemOpen
  \bibfield{author}{%
  \bibinfo {author} {\bibfnamefont{J.~H.}\ \bibnamefont{Chu}}, \bibinfo
  {author} {\bibfnamefont{J.~G.}\ \bibnamefont{Analytis}}, \bibinfo {author}
  {\bibfnamefont{C.}~\bibnamefont{Kucharczyk}},\ and\ \bibinfo {author}
  {\bibfnamefont{I.~R.}\ \bibnamefont{Fisher}},\ }%
  \bibfield{journal}{%
  \bibinfo {journal} {Phys. Rev. B}\ }%
  \textbf{\bibinfo {volume} {79}},\ \bibinfo {pages} {014506} (\bibinfo {year}
  {2009})%
  \bibAnnoteFile{NoStop}{Chu2009}%
\bibitem{Albenque2009}%
  \BibitemOpen
  \bibfield{author}{%
  \bibinfo {author} {\bibfnamefont{F.}~\bibnamefont{Rullier-Albenque}},
  \bibinfo {author} {\bibfnamefont{D.}~\bibnamefont{Colson}}, \bibinfo {author}
  {\bibfnamefont{A.}~\bibnamefont{Forget}},\ and\ \bibinfo {author}
  {\bibfnamefont{H.}~\bibnamefont{Alloul}},\ }%
  \bibfield{journal}{%
  \bibinfo {journal} {Phys. Rev. Lett.}\ }%
  \textbf{\bibinfo {volume} {103}},\ \bibinfo {pages} {057001} (\bibinfo {year}
  {2009})%
  \bibAnnoteFile{NoStop}{Albenque2009}%
\bibitem{Fang2009}%
  \BibitemOpen
  \bibfield{author}{%
  \bibinfo {author} {\bibfnamefont{L.}~\bibnamefont{Fang}}, \bibinfo {author}
  {\bibfnamefont{H.}~\bibnamefont{Luo}}, \bibinfo {author}
  {\bibfnamefont{P.}~\bibnamefont{Cheng}}, \bibinfo {author}
  {\bibfnamefont{Z.}~\bibnamefont{Wang}}, \bibinfo {author}
  {\bibfnamefont{Y.}~\bibnamefont{Jia}}, \bibinfo {author}
  {\bibfnamefont{G.}~\bibnamefont{Mu}}, \bibinfo {author}
  {\bibfnamefont{B.}~\bibnamefont{Shen}}, \bibinfo {author}
  {\bibfnamefont{I.~I.}\ \bibnamefont{Mazin}}, \bibinfo {author}
  {\bibfnamefont{L.}~\bibnamefont{Shan}}, \bibinfo {author}
  {\bibfnamefont{C.}~\bibnamefont{Ren}},\ and\ \bibinfo {author}
  {\bibfnamefont{H.~H.}\ \bibnamefont{Wen}},\ }%
  \bibfield{journal}{%
  \bibinfo {journal} {Phys. Rev. B}\ }%
  \textbf{\bibinfo {volume} {80}},\ \bibinfo {pages} {140508(R)} (\bibinfo
  {year} {2009})%
  \bibAnnoteFile{NoStop}{Fang2009}%
\bibitem{Boekelheide2007}%
  \BibitemOpen
  \bibfield{author}{%
  \bibinfo {author} {\bibfnamefont{Z.}~\bibnamefont{Boekelheide}}, \bibinfo
  {author} {\bibfnamefont{E.}~\bibnamefont{Helgren}},\ and\ \bibinfo {author}
  {\bibfnamefont{F.}~\bibnamefont{Hellman}},\ }%
  \bibfield{journal}{%
  \bibinfo {journal} {Phys. Rev. B}\ }%
  \textbf{\bibinfo {volume} {76}},\ \bibinfo {pages} {224429} (\bibinfo {year}
  {2007})%
  \bibAnnoteFile{NoStop}{Boekelheide2007}%
\bibitem{Dong2008}%
  \BibitemOpen
  \bibfield{author}{%
  \bibinfo {author} {\bibfnamefont{J.~K.}\ \bibnamefont{Dong}}, \bibinfo
  {author} {\bibfnamefont{L.}~\bibnamefont{Ding}}, \bibinfo {author}
  {\bibfnamefont{H.}~\bibnamefont{Wang}}, \bibinfo {author}
  {\bibfnamefont{X.~F.}\ \bibnamefont{Wang}}, \bibinfo {author}
  {\bibfnamefont{T.}~\bibnamefont{Wu}}, \bibinfo {author}
  {\bibfnamefont{G.}~\bibnamefont{Wu}}, \bibinfo {author}
  {\bibfnamefont{X.~H.}\ \bibnamefont{Chen}},\ and\ \bibinfo {author}
  {\bibfnamefont{S.~Y.}\ \bibnamefont{Li}},\ }%
  \bibfield{journal}{%
  \bibinfo {journal} {New J. Phys.}\ }%
  \textbf{\bibinfo {volume} {10}},\ \bibinfo {pages} {123031} (\bibinfo {year}
  {2008})%
  \bibAnnoteFile{NoStop}{Dong2008}%
\bibitem{Uchida1991}%
  \BibitemOpen
  \bibfield{author}{%
  \bibinfo {author} {\bibfnamefont{S.}~\bibnamefont{Uchida}}, \bibinfo {author}
  {\bibfnamefont{T.}~\bibnamefont{Ido}}, \bibinfo {author}
  {\bibfnamefont{H.}~\bibnamefont{Takagi}}, \bibinfo {author}
  {\bibfnamefont{T.}~\bibnamefont{Arima}}, \bibinfo {author}
  {\bibfnamefont{Y.}~\bibnamefont{Tokura}},\ and\ \bibinfo {author}
  {\bibfnamefont{S.}~\bibnamefont{Tajima}},\ }%
  \bibfield{journal}{%
  \bibinfo {journal} {Phys. Rev. B}\ }%
  \textbf{\bibinfo {volume} {43}},\ \bibinfo {pages} {7942} (\bibinfo {year}
  {1991})%
  \bibAnnoteFile{NoStop}{Uchida1991}%
\bibitem{Thomas1988}%
  \BibitemOpen
  \bibfield{author}{%
  \bibinfo {author} {\bibfnamefont{G.~A.}\ \bibnamefont{Thomas}}, \bibinfo
  {author} {\bibfnamefont{J.}~\bibnamefont{Orenstein}}, \bibinfo {author}
  {\bibfnamefont{D.~H.}\ \bibnamefont{Rapkine}}, \bibinfo {author}
  {\bibfnamefont{M.}~\bibnamefont{Capizzi}}, \bibinfo {author}
  {\bibfnamefont{A.~J.}\ \bibnamefont{Millis}}, \bibinfo {author}
  {\bibfnamefont{R.~N.}\ \bibnamefont{Bhatt}}, \bibinfo {author}
  {\bibfnamefont{L.~F.}\ \bibnamefont{Schneemeyer}},\ and\ \bibinfo {author}
  {\bibfnamefont{J.~V.}\ \bibnamefont{Waszczak}},\ }%
  \bibfield{journal}{%
  \bibinfo {journal} {Phys. Rev. Lett.}\ }%
  \textbf{\bibinfo {volume} {61}},\ \bibinfo {pages} {1313} (\bibinfo {year}
  {1988})%
  \bibAnnoteFile{NoStop}{Thomas1988}%
\bibitem{Nakajima2009}%
  \BibitemOpen
  \bibfield{author}{%
  \bibinfo {author} {\bibfnamefont{M.}~\bibnamefont{Nakajima}}, \bibinfo
  {author} {\bibfnamefont{S.}~\bibnamefont{Ishida}}, \bibinfo {author}
  {\bibfnamefont{K.}~\bibnamefont{Kihou}}, \bibinfo {author}
  {\bibfnamefont{Y.}~\bibnamefont{Tomioka}}, \bibinfo {author}
  {\bibfnamefont{T.}~\bibnamefont{Ito}}, \bibinfo {author}
  {\bibfnamefont{C.~H.}\ \bibnamefont{Lee}}, \bibinfo {author}
  {\bibfnamefont{H.}~\bibnamefont{Kito}}, \bibinfo {author}
  {\bibfnamefont{A.}~\bibnamefont{Iyo}}, \bibinfo {author}
  {\bibfnamefont{H.}~\bibnamefont{Eisaki}}, \bibinfo {author}
  {\bibfnamefont{K.~M.}\ \bibnamefont{Kojima}},\ and\ \bibinfo {author}
  {\bibfnamefont{S.}~\bibnamefont{Uchida}},\ }%
  \bibfield{journal}{%
  \bibinfo {journal} {Physica C}}%
   (\bibinfo {year} {to be published}),\ \bibinfo {note}
  {doi:10.1016/j.physc.2009.11.071}%
  \bibAnnoteFile{NoStop}{Nakajima2009}%
\bibitem{Ishida2009}%
  \BibitemOpen
  \bibfield{author}{%
  \bibinfo {author} {\bibfnamefont{S.}~\bibnamefont{Ishida}}, \bibinfo {author}
  {\bibfnamefont{M.}~\bibnamefont{Nakajima}}, \bibinfo {author}
  {\bibfnamefont{Y.}~\bibnamefont{Yomioka}}, \bibinfo {author}
  {\bibfnamefont{T.}~\bibnamefont{Ito}}, \bibinfo {author}
  {\bibfnamefont{K.}~\bibnamefont{Miyazawa}}, \bibinfo {author}
  {\bibfnamefont{H.}~\bibnamefont{Kito}}, \bibinfo {author}
  {\bibfnamefont{C.~H.}\ \bibnamefont{Lee}}, \bibinfo {author}
  {\bibfnamefont{M.}~\bibnamefont{Ishikado}}, \bibinfo {author}
  {\bibfnamefont{S.}~\bibnamefont{Shamoto}}, \bibinfo {author}
  {\bibfnamefont{A.}~\bibnamefont{Iyo}}, \bibinfo {author}
  {\bibfnamefont{H.}~\bibnamefont{Eisaki}}, \bibinfo {author}
  {\bibfnamefont{K.~M.}\ \bibnamefont{Kojima}},\ and\ \bibinfo {author}
  {\bibfnamefont{S.}~\bibnamefont{Uchida}},\ }%
  \bibfield{journal}{%
  \bibinfo {journal} {Phys. Rev. B}\ }%
  \textbf{\bibinfo {volume} {81}},\ \bibinfo {pages} {094515} (\bibinfo {year}
  {2010})%
  \bibAnnoteFile{NoStop}{Ishida2009}%
\bibitem{Miyazawa2009}%
  \BibitemOpen
  \bibfield{author}{%
  \bibinfo {author} {\bibfnamefont{K.}~\bibnamefont{Miyazawa}}, \bibinfo
  {author} {\bibfnamefont{K.}~\bibnamefont{Kihou}}, \bibinfo {author}
  {\bibfnamefont{P.~M.}\ \bibnamefont{Shirage}}, \bibinfo {author}
  {\bibfnamefont{C.~H.}\ \bibnamefont{Lee}}, \bibinfo {author}
  {\bibfnamefont{H.}~\bibnamefont{Kito}}, \bibinfo {author}
  {\bibfnamefont{H.}~\bibnamefont{Eisaki}},\ and\ \bibinfo {author}
  {\bibfnamefont{A.}~\bibnamefont{Iyo}},\ }%
  \bibfield{journal}{%
  \bibinfo {journal} {J. Phys. Soc. Jpn.}\ }%
  \textbf{\bibinfo {volume} {78}},\ \bibinfo {pages} {034712} (\bibinfo {year}
  {2009})%
  \bibAnnoteFile{NoStop}{Miyazawa2009}%
\bibitem{Wu2009b}%
  \BibitemOpen
  \bibfield{author}{%
  \bibinfo {author} {\bibfnamefont{D.}~\bibnamefont{Wu}}, \bibinfo {author}
  {\bibfnamefont{N.}~\bibnamefont{Bari\v{s}i\'{c}}}, \bibinfo {author}
  {\bibfnamefont{P.}~\bibnamefont{Kallina}}, \bibinfo {author}
  {\bibfnamefont{A.}~\bibnamefont{Faridian}}, \bibinfo {author}
  {\bibfnamefont{B.}~\bibnamefont{Gorchunov}}, \bibinfo {author}
  {\bibfnamefont{N.}~\bibnamefont{Drichko}}, \bibinfo {author}
  {\bibfnamefont{L.~J.}\ \bibnamefont{Li}}, \bibinfo {author}
  {\bibfnamefont{X.}~\bibnamefont{Lin}}, \bibinfo {author}
  {\bibfnamefont{G.~H.}\ \bibnamefont{Cao}}, \bibinfo {author}
  {\bibfnamefont{Z.~A.}\ \bibnamefont{Xu}}, \bibinfo {author}
  {\bibfnamefont{N.~L.}\ \bibnamefont{Wang}},\ and\ \bibinfo {author}
  {\bibfnamefont{M.}~\bibnamefont{Dressel}},\ }%
  \bibinfo {note} {{arXiv:0912.3334 (unpublished). In comparison with the
  present result, the $\omega=0$ peak of the optical conductivity spectrum of
  \BFA{} (also that in Ref.~\onlinecite{Hu2008}) is anomalously high, even
  compared with their own results on SrFe$_2$As$_2$ and EuFe$_2$As$_2$.}}%
  \bibAnnoteFile{Stop}{Wu2009b}%
\bibitem{Terashima2009}%
  \BibitemOpen
  \bibfield{author}{%
  \bibinfo {author} {\bibfnamefont{K.}~\bibnamefont{Terashima}}, \bibinfo
  {author} {\bibfnamefont{Y.}~\bibnamefont{Sekiba}}, \bibinfo {author}
  {\bibfnamefont{J.~H.}\ \bibnamefont{Bowen}}, \bibinfo {author}
  {\bibfnamefont{K.}~\bibnamefont{Nakayama}}, \bibinfo {author}
  {\bibfnamefont{T.}~\bibnamefont{Kawabata}}, \bibinfo {author}
  {\bibfnamefont{T.}~\bibnamefont{Sato}}, \bibinfo {author}
  {\bibfnamefont{P.}~\bibnamefont{Richard}}, \bibinfo {author}
  {\bibfnamefont{Y.-M.}\ \bibnamefont{Xu}}, \bibinfo {author}
  {\bibfnamefont{L.~J.}\ \bibnamefont{Li}}, \bibinfo {author}
  {\bibfnamefont{G.~H.}\ \bibnamefont{Cao}}, \bibinfo {author}
  {\bibfnamefont{Z.-A.}\ \bibnamefont{Xu}}, \bibinfo {author}
  {\bibfnamefont{H.}~\bibnamefont{Ding}},\ and\ \bibinfo {author}
  {\bibfnamefont{T.}~\bibnamefont{Takahashi}},\ }%
  \bibfield{journal}{%
  \bibinfo {journal} {Proc. Natl. Acad. Sci.}\ }%
  \textbf{\bibinfo {volume} {106}},\ \bibinfo {pages} {7330} (\bibinfo {year}
  {2009})%
  \bibAnnoteFile{NoStop}{Terashima2009}%
\bibitem{Kim2009}%
  \BibitemOpen
  \bibfield{author}{%
  \bibinfo {author} {\bibfnamefont{K.~W.}\ \bibnamefont{Kim}}, \bibinfo
  {author} {\bibfnamefont{M.}~\bibnamefont{R{\"{o}}ssle}}, \bibinfo {author}
  {\bibfnamefont{A.}~\bibnamefont{Dubroka}}, \bibinfo {author}
  {\bibfnamefont{V.~K.}\ \bibnamefont{Malik}}, \bibinfo {author}
  {\bibfnamefont{T.}~\bibnamefont{Wolf}},\ and\ \bibinfo {author}
  {\bibfnamefont{C.}~\bibnamefont{Bernhard}},\ }%
  \bibinfo {note} {arXiv:0912.0140 (unpublished)}%
  \bibAnnoteFile{NoStop}{Kim2009}%
\bibitem{Heumen2009}%
  \BibitemOpen
  \bibfield{author}{%
  \bibinfo {author} {\bibfnamefont{E.}~\bibnamefont{van Heumen}}, \bibinfo
  {author} {\bibfnamefont{Y.}~\bibnamefont{Huang}}, \bibinfo {author}
  {\bibfnamefont{S.}~\bibnamefont{de~Jong}}, \bibinfo {author}
  {\bibfnamefont{A.~B.}\ \bibnamefont{Kuzmenko}}, \bibinfo {author}
  {\bibfnamefont{M.~S.}\ \bibnamefont{Golden}},\ and\ \bibinfo {author}
  {\bibfnamefont{D.}~\bibnamefont{van~der Marel}},\ }%
  \bibinfo {note} {arXiv:0912.0636 (unpublished)}%
  \bibAnnoteFile{NoStop}{Heumen2009}%
\bibitem{Gorshunov2009}%
  \BibitemOpen
  \bibfield{author}{%
  \bibinfo {author} {\bibfnamefont{B.}~\bibnamefont{Gorshunov}}, \bibinfo
  {author} {\bibfnamefont{D.}~\bibnamefont{Wu}}, \bibinfo {author}
  {\bibfnamefont{A.~A.}\ \bibnamefont{Voronkov}}, \bibinfo {author}
  {\bibfnamefont{P.}~\bibnamefont{Kallina}}, \bibinfo {author}
  {\bibfnamefont{K.}~\bibnamefont{Iida}}, \bibinfo {author}
  {\bibfnamefont{S.}~\bibnamefont{Haindl}}, \bibinfo {author}
  {\bibfnamefont{F.}~\bibnamefont{Kurth}}, \bibinfo {author}
  {\bibfnamefont{L.}~\bibnamefont{Schultz}}, \bibinfo {author}
  {\bibfnamefont{B.}~\bibnamefont{Holzapfel}},\ and\ \bibinfo {author}
  {\bibfnamefont{M.}~\bibnamefont{Dressel}},\ }%
  \bibfield{journal}{%
  \bibinfo {journal} {Phys. Rev. B}\ }%
  \textbf{\bibinfo {volume} {81}},\ \bibinfo {pages} {060509(R)} (\bibinfo
  {year} {2010})%
  \bibAnnoteFile{NoStop}{Gorshunov2009}%
\bibitem{Samuely2009}%
  \BibitemOpen
  \bibfield{author}{%
  \bibinfo {author} {\bibfnamefont{P.}~\bibnamefont{Samuely}}, \bibinfo
  {author} {\bibfnamefont{Z.}~\bibnamefont{Pribulov\'{a}}}, \bibinfo {author}
  {\bibfnamefont{P.}~\bibnamefont{Szab\'{o}}}, \bibinfo {author}
  {\bibfnamefont{G.}~\bibnamefont{Prist\'{a}\v{s}}}, \bibinfo {author}
  {\bibfnamefont{S.~L.}\ \bibnamefont{Bud'ko}},\ and\ \bibinfo {author}
  {\bibfnamefont{P.~C.}\ \bibnamefont{Canfield}},\ }%
  \bibfield{journal}{%
  \bibinfo {journal} {Physica C}\ }%
  \textbf{\bibinfo {volume} {469}},\ \bibinfo {pages} {507} (\bibinfo {year}
  {2009})%
  \bibAnnoteFile{NoStop}{Samuely2009}%
\bibitem{Williams2009}%
  \BibitemOpen
  \bibfield{author}{%
  \bibinfo {author} {\bibfnamefont{T.~J.}\ \bibnamefont{Williams}}, \bibinfo
  {author} {\bibfnamefont{A.~A.}\ \bibnamefont{Aczel}}, \bibinfo {author}
  {\bibfnamefont{E.}~\bibnamefont{Baggio-Saitovitch}}, \bibinfo {author}
  {\bibfnamefont{S.~L.}\ \bibnamefont{Bud'ko}}, \bibinfo {author}
  {\bibfnamefont{P.~C.}\ \bibnamefont{Canfield}}, \bibinfo {author}
  {\bibfnamefont{J.~P.}\ \bibnamefont{Carlo}}, \bibinfo {author}
  {\bibfnamefont{T.}~\bibnamefont{Goko}}, \bibinfo {author}
  {\bibfnamefont{J.}~\bibnamefont{Munevar}}, \bibinfo {author}
  {\bibfnamefont{N.}~\bibnamefont{Ni}}, \bibinfo {author}
  {\bibfnamefont{Y.~J.}\ \bibnamefont{Uemura}}, \bibinfo {author}
  {\bibfnamefont{W.}~\bibnamefont{Yu}},\ and\ \bibinfo {author}
  {\bibfnamefont{G.~M.}\ \bibnamefont{Luke}},\ }%
  \bibfield{journal}{%
  \bibinfo {journal} {Phys. Rev. B}\ }%
  \textbf{\bibinfo {volume} {80}},\ \bibinfo {pages} {094501} (\bibinfo {year}
  {2009})%
  \bibAnnoteFile{NoStop}{Williams2009}%
\bibitem{Gordon2009}%
  \BibitemOpen
  \bibfield{author}{%
  \bibinfo {author} {\bibfnamefont{R.~T.}\ \bibnamefont{Gordon}}, \bibinfo
  {author} {\bibfnamefont{N.}~\bibnamefont{Ni}}, \bibinfo {author}
  {\bibfnamefont{C.}~\bibnamefont{Martin}}, \bibinfo {author}
  {\bibfnamefont{M.~A.}\ \bibnamefont{Tanatar}}, \bibinfo {author}
  {\bibfnamefont{M.~D.}\ \bibnamefont{Vannette}}, \bibinfo {author}
  {\bibfnamefont{H.}~\bibnamefont{Kim}}, \bibinfo {author}
  {\bibfnamefont{G.~D.}\ \bibnamefont{Samolyuk}}, \bibinfo {author}
  {\bibfnamefont{J.}~\bibnamefont{Schmalian}}, \bibinfo {author}
  {\bibfnamefont{S.}~\bibnamefont{Nandi}}, \bibinfo {author}
  {\bibfnamefont{A.}~\bibnamefont{Kreyssig}}, \bibinfo {author}
  {\bibfnamefont{A.~I.}\ \bibnamefont{Goldman}}, \bibinfo {author}
  {\bibfnamefont{J.~Q.}\ \bibnamefont{Yan}}, \bibinfo {author}
  {\bibfnamefont{S.~L.}\ \bibnamefont{Bud'ko}}, \bibinfo {author}
  {\bibfnamefont{P.~C.}\ \bibnamefont{Canfield}},\ and\ \bibinfo {author}
  {\bibfnamefont{R.}~\bibnamefont{Prozorov}},\ }%
  \bibfield{journal}{%
  \bibinfo {journal} {Phys. Rev. Lett.}\ }%
  \textbf{\bibinfo {volume} {102}},\ \bibinfo {pages} {127004} (\bibinfo {year}
  {2009})%
  \bibAnnoteFile{NoStop}{Gordon2009}%
\bibitem{Canfield2009}%
  \BibitemOpen
  \bibfield{author}{%
  \bibinfo {author} {\bibfnamefont{P.~C.}\ \bibnamefont{Canfield}}, \bibinfo
  {author} {\bibfnamefont{S.~L.}\ \bibnamefont{Bud'ko}}, \bibinfo {author}
  {\bibfnamefont{N.}~\bibnamefont{Ni}}, \bibinfo {author}
  {\bibfnamefont{J.~Q.}\ \bibnamefont{Yan}},\ and\ \bibinfo {author}
  {\bibfnamefont{A.}~\bibnamefont{Kracher}},\ }%
  \bibfield{journal}{%
  \bibinfo {journal} {Phys. Rev. B}\ }%
  \textbf{\bibinfo {volume} {80}},\ \bibinfo {pages} {060501(R)} (\bibinfo
  {year} {2009})%
  \bibAnnoteFile{NoStop}{Canfield2009}%
\bibitem{Chu2010}%
  \BibitemOpen
  \bibfield{author}{%
  \bibinfo {author} {\bibfnamefont{J.-H.}\ \bibnamefont{Chu}}, \bibinfo
  {author} {\bibfnamefont{J.~G.}\ \bibnamefont{Analytis}}, \bibinfo {author}
  {\bibfnamefont{K.~D.}\ \bibnamefont{Greve}}, \bibinfo {author}
  {\bibfnamefont{P.~L.}\ \bibnamefont{McMahon}}, \bibinfo {author}
  {\bibfnamefont{Z.}~\bibnamefont{Islam}}, \bibinfo {author}
  {\bibfnamefont{Y.}~\bibnamefont{Yamamoto}},\ and\ \bibinfo {author}
  {\bibfnamefont{I.~R.}\ \bibnamefont{Fisher}},\ }%
  \bibinfo {note} {arXiv:1002.3364 (unpublished)}%
  \bibAnnoteFile{NoStop}{Chu2010}%
\end{thebibliography}%

\end{document}